# Living Globe: Tridimensional interactive visualization of world demographic data

Eduardo Duarte[,1], Pedro Bordonhos[,1], Paulo Dias[1,2], Beatriz Sousa Santos[1,2]

[1]Department of Electronics Telecommunications and Informatics, Univ. Aveiro, Portugal
[2]Institute of Electronics and Informatics Engineering of Aveiro/IEETA, Portugal

`emod@ua.pt, bordonhos@ua.pt, paulo.dias@ua.pt, bss@ua.pt`

**Abstract.**
This paper presents Living Globe, an application for visualization of demographic data supporting the temporal comparison of data from several countries represented on a 3D globe. Living Globe allows the visual exploration of the following demographic data: total population, population density and growth, crude birth and death rates, life expectancy, net migration and population percentage of different age groups. While offering unexperienced users a default mapping of these data variables into visual variables, Living Globe allows more advanced users to select the mapping, increasing its flexibility.
The main aspects of the Living Globe model and prototype are described as well as the evaluation results obtained using heuristic evaluation and usability testing. Some conclusions and ideas for future work are also presented.

**Keywords:** Information Visualization, demographic data, 3D globe, WebGL Globe, usability evaluation, heuristic evaluation.

## 1 Introduction

This paper presents Living Globe, a 3D web visualization application meant to support the study of demographic data allowing users to compare data corresponding to several countries along the years. Living Globe allows the visual exploration of demographic data represented on a 3D globe and offers functionality not available in other demographic data visualization applications.
While 3D data visualization may have advantages and disadvantages regarding 2D solutions [1], and may be more appropriate in specific contexts, we consider the visualization of demographic data on a 3D globe to potentially be a more intuitive and useful approach. There are already other applications allowing the 3D visualization of demographic data on a globe mapping total population or population density (for example) into the height of bars positioned on the globe at the corresponding location. "World Population" [2] and "China and US Population" [3] are two such applications, both based on WebGL Globe [4]; however, the applications we have found and ana-lyzed are very limited regarding the represented demographic data, its interactivity and its usability. Throughout this paper we describe the main aspects of our proposal in section 2, the prototype built to test it in section 3, and the results obtained with heuristic evaluation and tests involving users performed to test usability in section 4. Finally, conclusions and ideas for future work are presented in section 5.

## 2  Related Work

In the following section we present 3D data web applications that allow data visualization on a globe and provide interaction to some extent which have inspired our proposal (Fig.1). All except one allow visualizing total population or population density mapped as vertical bars with variable height on a globe.

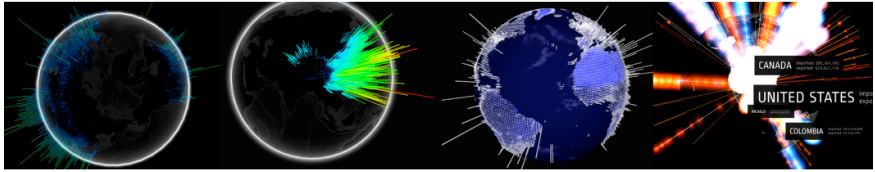

**Fig. 1.** Related Visualizations (from left to right): 'WebGL Globe - World Population', 'WebGL Globe - China and US Population', 'World Population Density - 2010' and 'Small Arms and Ammunition'.

These applications were implemented using the WebGL technology or other APIs allowing 3D representations in Javascript, like three.js [5]. These technologies allow the creation of interactive 3D visual objects with textures and shaders in a canvas, and make the solution compatible with various types of devices, requiring only internet access and a browser supporting these technologies (such as the most currently used browsers). One such platform is the WebGL Globe developed in the scope of the "Chrome Experiments" project based on the native API of WebGL, which is an open platform that allows the visualization of any set of spatial data in a tridimensional globe. Using this API, a user can easily map any data to available graphical elements (a bar by default) and positioned in an interactive 3D globe.
As mentioned, two projects featured at the WebGL Globe site allow visualizing demographic data: "World Population" [2] and "China and US Population"[3]. While both represent data provided by the Socioeconomic Data and Application Center (SEDAC), "China and US Population" represents a more limited data set; "World Population Density" [6] is another example that presents SEDAC demographic data on a globe, implemented in three.js. In these three examples the population of all countries is represented by the height of vertical bars. While the first two offer as single functionality, the possibility of rotating the globe, the latter is not interactive, since the globe rotates continuously which implies that the user has to wait until the globe assumes an adequate position to visualize the population of a specific country.

Another example that has inspired us in spite of visualizing non-demographic data is "Small Arms and Ammunition - Imports & Exports" [7]. It is an interactive visualization of government-authorized small arms and ammunition trades from 1992 to 2000. The year is selected through a slider and, unlike the previous examples, this visualization supports selection and search with automatic completion of countries. The traffic between any two countries is represented by curved lines connecting the two, taking advantage of the third dimension to prevent it from occulting countries and the selection feedback.

The analyzed demographic data visualization applications have limitations regarding interaction when compared to "Small Arms and Ammunition", which features searching and selection functionality that might be easily integrated in a demographic data application resulting in a higher usability. Moreover, the tri-dimensionality of the user interface might be further explored. These ideas were incorporated in our proposal, the Living Globe, described in the next section.

## 3 Living Globe: the proposal

Living Globe allows the visual exploration of the following demographic data along a set period of time: total population, population density and growth, crude birth and death rates, life expectancy, net migration and population percentage of different age groups. It is targeted to users that have some computer and statistics literacy. While offering unexperienced users a default mapping of these data variables into visual variables, Living Globe allows more advanced users to select the mapping they intent to use. This means that these users have the possibility of acting upon an earlier stage of the visualization reference model [8] making Living Globe a more flexible tool. In order to support this feature, three visual variables may be selected to map a data variable: i) height of vertical bars (directly proportional to the data value) ii) color of vertical bars (in a color scale ranging from blue to yellow) and iii) color of the countries on the globe (in a scale ranging from red to green). An adequate selection of the data variables and their mapping to the visual variables may help the identification and study of potential relations among data variables.

According to Robertson et al. [9], it is likely that using 3D visualization can "maximize effective use of screen space", as it enables the simultaneous representation (in a single view) of a larger part of the data (even if it implies distortion). These authors also deem that interactive animation supports object constancy; for instance, when the user rotates a representation of complex data, it will be easier to remember the relationships of what is visualized. Cognitive load is shifted to the perceptual system, which frees users' cognitive capacity to perform tasks with the application.

After the user customizes the mapping of the variables to analyse (or simply uses the default mapping), Living Globe allows the visualization of the selected demographic data on an interactive 3D globe (users can freely rotate the globe) while offering the following filtering functionality: i) selection of the country to analyze; ii) selection of the year to analyze; iii) definition of minimum and maximum data values that should be visualized; iv) search with dynamic suggestion of the country names. The time selection provides support for analysis of temporal evolution of data; the value interval selection is important, for instance, when visualizing the population of small countries (eg. Portugal) leaving out countries with large population (e.g. China). The dynamic search allows users to look more efficiently for countries when they do not know their location. Moreover, to ease the analysis of temporal evolution of data variables mapped to the three visual variables, it is possible to maintain configured limits to filter each variable when the selected year changes. Also, some variables can be displayed numerically on the user interface using country selection, supporting a more accurate analysis of the data. All of this functionality may be picked by the user according to the task at hand and instantiate to some extent Shneiderman's "Visualization seeking mantra" [10], namely, overview first, filter and details on demand.

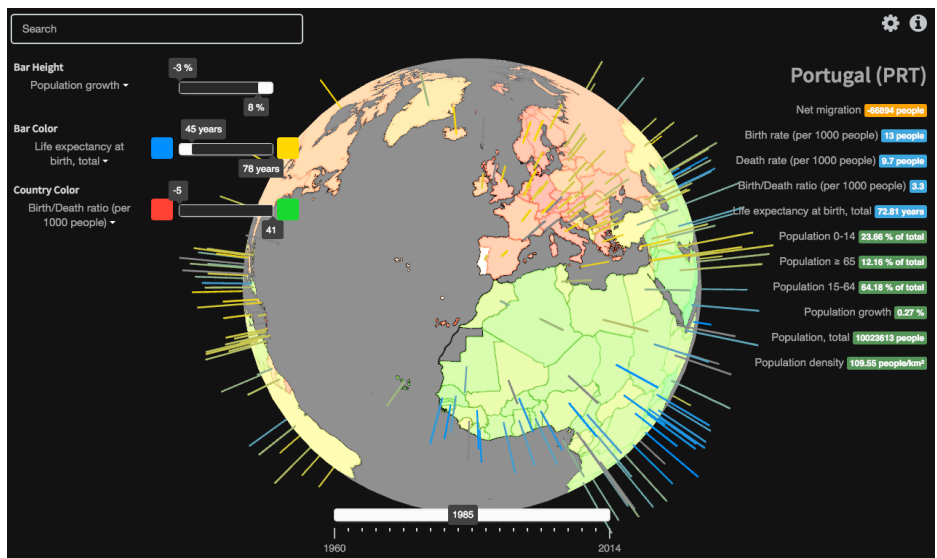

**Fig. 2.** Living Globe showing demographic data for Portugal in 1985

Fig. 2 shows the main aspects of the Living Globe user interface (UI), featuring the 3D interactive globe at the center, and various widgets allowing the selection of the above mentioned functionality: a search box for countries (upper left), sliders to set the filtering limits of the three variables mapped to the visual variables on the globe (middle left), numerical data corresponding to the selected country (top right), and a slider to select the year (beneath the globe).

Fig. 3 illustrates the filtering functionality regarding the data value interval to visualize as the countries color. On the left all the countries are visualized and most countries are represented with similar colors; on the right countries with large population (as China and India) are filtered out resulting in a visualization of the remaining countries with much more diverse and distinguishable colors.

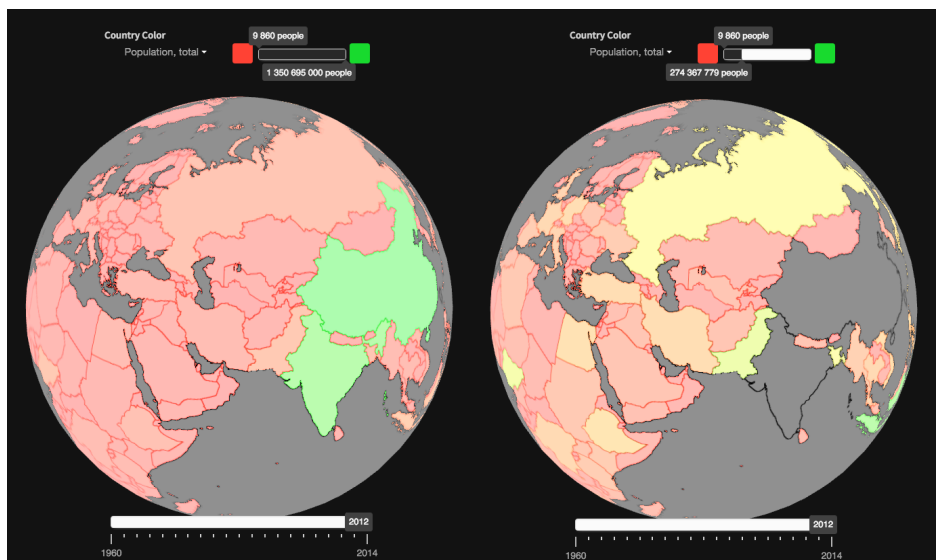

**Fig. 3.** Living Globe showing total population mapped to the country color: the data from all countries is displayed on the left, and filtering out large population countries (such as China and India) on the right (resulting in a representation with much more diverse and distinguishing colors for the remaining countries).

## 4   Implementing the Prototype

To test the Living Globe proposal a prototype was developed allowing the visualization of data corresponding to the period between 1980 and 2014, obtained from the World Bank [11] with an open license for non-commercial usage.

The prototype (whose code is open-source and available at [12]), was built using web technologies such as HTML, CSS and Javascript, as well as some Javascript libraries such as three.js, chroma.js [13] and jQuery [14], and was successfully tested for the following browsers: Firefox 44.0.2, Chrome 48.0.2564.109 (64-bit) and Safari 9.0.3. In this section some details concerning the implementation of the prototype are described.

### 4.1   Data

Data is structured in a table where each country-indicator corresponds to a line and a column corresponds to a year. The net migration indicator is registered only every five years. Countries are identified by their official name as well as by their ISO 3166 alpha-3 code, a three letter code identifying univocally each country [15]. The geographic location of the countries used to create the vertical bar representations are characterized by latitude-longitude pairs obtained from the Sokrata Open Data Portal [16] a portal providing data for non-commercial purposes also in tabular format. All data is collected asynchronously using Ajax technology and analyzed to obtain the data variables minimum and maximum values, which are used for data normalization.

### 4.2 Globe and countries

The country selection and their coloring according to a variable implies the identification of and drawing over country areas on a tridimensional object. As a first solution for this challenge, a CSV (Comma Separated Values) file containing a set of latitude-longitude pairs defining the polygonal border of each country was used [17]. To identify a selected country the cursor location was converted to latitude-longitude and compared with the country borders; however, this implementation was not efficient as drawing multiple polygons (whose properties change whenever the year or the filtering settings change) was resource-heavy and negatively affected the user experience.

A second solution, which is less computationally expensive, yielding a better user experience, uses textures and a raycaster, and was inspired by the "Small Arms and Ammunition" example mentioned in section 2. Five different textures are superimposed on the globe (Fig. 4): a) blend, using an image (Fig. 4A) with a gradient effect over the sea area for aesthetical purposes; b) outline, using an image (Fig. 4B) with lines over the countries frontiers, allowing a better differentiation of different countries when these are displaying the same color; c) lookup, using an image (Fig. 4C) that colors country territories with a unique luminosity level of grey; d) ratio, initially transparent and where country territories are colored with the color corresponding to the value of the mapped data variable; and e) select, initially transparent and where country territories are colored in white to provide feedback corresponding to the selected country.

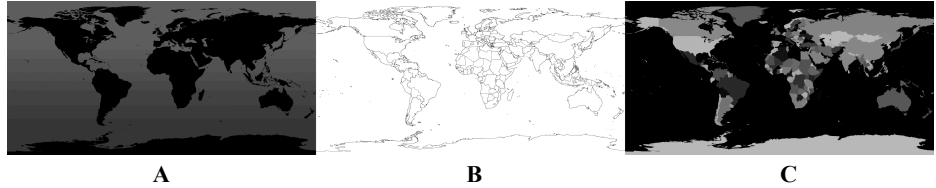

**A**            **B**            **C**

**Fig. 4.** Images used as textures in Living Globe: blend, outline and lookup (from left to right)

A manually generated JSON file mapping grey levels to ISO 3166-1 alpha-3 codes was used. Since the demographic data used the same code per country, it was easy to associate them to each grey color.

A color scale obtained from a gradient between red and green (representing the minimum and the maximum values respectively) was created using the chroma.js library to represent the normalized value ($V_{norm}$) of the data variable mapped to the country color, according to the mathematical expression 1, where dv is the data variable to represent, c is the country and y is the selected year.

$$V_{norm}(dv, c, y) = (V(dv, c, y) - Min(dv, y)) / (Max(dv, y) - Min(dv, y)) \qquad (1)$$

### 4.3 Vertical bars

A similar procedure was used to map another variable to the color of the vertical bars. A gradient between blue and yellow (representing the minimum and the maximum values respectively) was used for this mapping.

Finally, the height of the bars (mapping a third data variable) is also obtained using expression 1 and representing the maximum data value with a height of 100 pixels, which was found adequate taking into consideration the initial camera position and the globe scale.

## 5 Usability evaluation

This section presents the main results obtained using an evaluation approach that we have been employing in our previous work [18] using two widely known and applied usability evaluation methods: heuristic evaluation [19], and usability testing used in order to improve the prototype (i.e. as a formative evaluation) [20]. The heuristic evaluation was performed using the Nielsen's usability heuristics by two evaluators, and three different sets of heuristics by one evaluator, all having some experience in using the method to analyze Information Visualization applications. During the usability test four users were observed while performing five tasks with different degrees of complexity, answered a post-task questionnaire, and gave some feedback concerning possible improvements. The most relevant findings are discussed in the following sections.

### 5.1 Heuristic Evaluation

Heuristic evaluation is a usability evaluation method extensively used to find potential usability problems proposed by Nielsen [19,21]. It is performed by examining a user interface taking into consideration a set of heuristics that should be complied with. A list of problems corresponding to noncompliant aspects, categorized according to their estimated impact on user performance or acceptance and rated according to their severity is produced. This list is supposed to be used by the development team to prioritize the fixes to improve the user interface.

According to Tory and Möller [22,23] heuristic evaluation is a useful expert review method to evaluate visualization applications and they recommended the use of visualization-specific heuristics. Several visualization-oriented heuristic sets have been proposed and used to evaluate visualization applications [24]. We have used the Nielsen's heuristics, a general heuristics set and two visualization-specific sets of heuristics [25,26]. Several positive aspects were found as well as several minor potential usability problems and four issues with a severity equal or higher than 3 (in a scale of 1-5, where 5 is a catastrophic usability problem). Fig. 5 illustrates one such problem related with the perception of the color used to represent visually a variable as the color of the country on the globe: the different colors of the scale used may be difficult to discriminate by color blind people. This can be alleviated by offering alternative color scales that the user can pick for both the bar colors and the country colors.

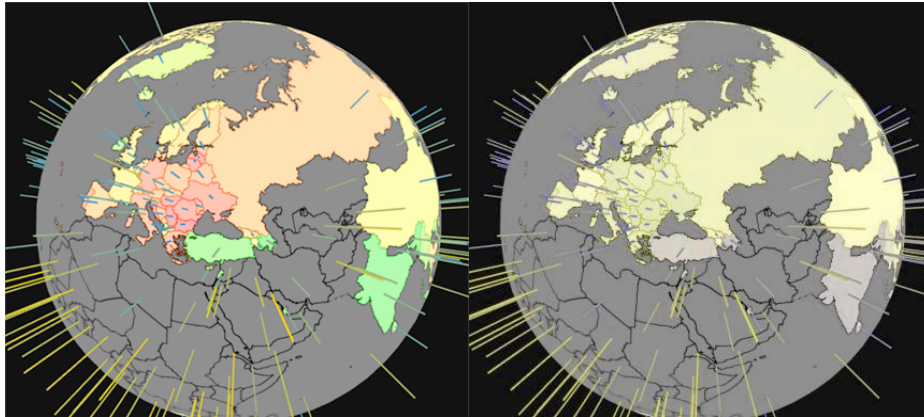

**Fig. 5.** Living Globe as seen by normal observers on the left, and color-blind observers (with Green-Blind/Deuteranopia) one the right, simulated by the Coblis simulator [27].

The remaining potential usability problems found having a higher severity grade were related to a deficient spatial organization that may occur when the browser window does not have enough size to accommodate all the visual elements of Living Globe and several elements are superimposed, and when the filter sliders are so close to each other that the numerical value on the right starts to occlude the values on the left (Fig. 6A). To lessen the first issue, the widgets layout was slightly changed, and to solve the latter issue, the numerical values were displayed on different sides of the sliders, one over and one under (Fig. 6B).

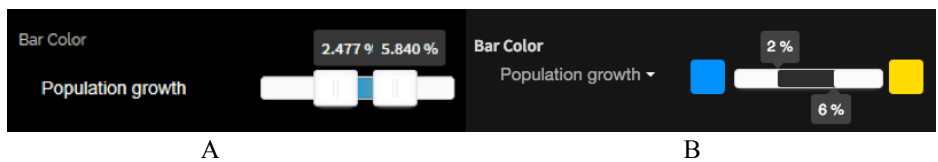

A                              B

**Fig. 6.** Usability issue: version A has a slider whose numerical value on the right partially occludes the value on the left. Version B is the solution found for the latest version of Living Globe.

Most of the potential usability problems detected by the evaluators have been corrected or alleviated in the subsequent versions of Living Globe.

### 5.2 Usability test

A simple usability test was achieved with a small set of tasks to be performed by participants while observed by the experimenter, a questionnaire, and a short interview to obtain participants' feedback concerning the application. Before executing these tasks, the participants where allowed to test the visualization at will and without any time limit in order to experiment with all of its features.

The questionnaire included questions regarding the difficulty and confidence in performing each task, and the overall satisfaction with the application.

The tasks corresponded to answering the following five questions:
1. What is the total population of Portugal?
2. What is the population growth of Benin?
3. Which country has the highest crude birth rate (without using search or selection)?
4. Which country has the highest birth/death ratio in the year 1990: Argentina or Brazil (without using search or selection)?
5. What is the country having the highest crude death rate value (representing this variable as country color and without using search or selection)?

The first three tasks are relatively easy and their main goal was to let the participants become more familiar with the application while encouraging them to explore the available functionalities (such as search). The last two tasks are more complex and were included in the test to find if they are understandable by the users and usable enough.
The usability test was performed with the collaboration of four participants, two having low computer literacy and no experience with information visualization applications, aged 23 and 59, and two being students at the Masters level in the Computing field, with high computer literacy but no experience with information visualization evaluation procedures, both aged 23.
In general, all tests lasted approximately 7 minutes and all participants completed all tasks without requiring assistance. The fourth and fifth tasks were answered with less confidence, and two of the participants completed task number five incorrectly.
Additionally, all users preferred to observe the data variables using country colors and text over bar colors and bar heights, which leads us to believe that vertical bars are not an appropriate visual variable to represent spatial data when compared to area coloring and text.
Along with the tasks, users also tested and provided feedback on a few variants of the visualization with additional widgets, like the inclusion of box-plots that represent the minimum, maximum and median values of the data variables before and after filtering, and the possibility of showing the numerical data in a tooltip that follows the mouse cursor instead of being placed statically in the top right corner. For the first variant, none of the users found the box-plots particularly useful. For the second variant, user preferences were split between static and tooltip.

Observation and participants' feedback suggest that the application was easy to learn and use, the filtering functionality was very useful, but the color scale used to colorize the countries might be misinterpreted (as green and red may have different cultural meanings). Once again, the addition of alternative color scales could mitigate this problem since the users could pick a more appropriate color scheme to represent the displayed data variables.

## 6      Conclusions and future work

This paper describes a 3D interactive visualization tool that allows users to visually explore demographic data represented on a globe. The general outcome of the evaluation is that Living Globe is an interesting tool to visualize demographic data and has potential to become more useful. The evaluation pointed out some negative points that were corrected in future versions, such as allowing the user selection of custom color scales, and provided some ideas for future work, such as the automatic adjustment of the widgets to the data characteristics (e.g. if the data does not allow the selection of a time period, the corresponding slider should not be present in the interface). Moreover, the conceptual model of Living Globe may be used beyond demographic data visualization and we consider it as a step to a more general API, allowing interactive representation of any kind of spatial data on a globe.

**Acknowledgments.** The authors are grateful to Beatriz Quintino Ferreira, Bruno Andrade, Bruno Garcia, Isabel Nascimento, Luís Afonso, Luís Silva, Pedro Miguel, Rui Simões e João Pedrosa for their valuable contribution to the usability evaluation of the prototype. This work was partially funded by National Funds through FCT - Foundation for Science and Technology, in the context of the projects UID/CEC/00127/2013 and Incentivo/EEI/UI0127/2014.